\documentclass[twoside,twocolumn,9pt]{article}
\usepackage{extsizes}
\usepackage[super,sort&compress,comma]{natbib} 
\usepackage[version=3]{mhchem}
\usepackage[left=1.5cm, right=1.5cm, top=1.785cm, bottom=2.0cm]{geometry}
\usepackage{balance}
\usepackage{mathptmx}
\usepackage{sectsty}
\usepackage{graphicx} 
\usepackage{lastpage}
\usepackage[format=plain,justification=justified,singlelinecheck=false,font={stretch=1.125,small,sf},labelfont=bf,labelsep=space]{caption}
\usepackage{float}
\usepackage{fancyhdr}
\usepackage{fnpos}
\usepackage[english]{babel}
\addto{\captionsenglish}{%
  \renewcommand{\refname}{Notes and references}
}
\usepackage{array}
\usepackage{droidsans}
\usepackage{charter}
\usepackage[T1]{fontenc}
\usepackage[usenames,dvipsnames]{xcolor}
\usepackage{setspace}
\usepackage[compact]{titlesec}
\usepackage{hyperref}

\usepackage{amsmath,amssymb}
\usepackage{widetext}
\usepackage{bm}
\usepackage{stfloats}

\usepackage{epstopdf}

\definecolor{cream}{RGB}{222,217,201}

\begin{document}

\pagestyle{fancy}
\thispagestyle{plain}
\fancypagestyle{plain}{
\renewcommand{\headrulewidth}{0pt}
}

\makeFNbottom
\makeatletter
\renewcommand\LARGE{\@setfontsize\LARGE{15pt}{17}}
\renewcommand\Large{\@setfontsize\Large{12pt}{14}}
\renewcommand\large{\@setfontsize\large{10pt}{12}}
\renewcommand\footnotesize{\@setfontsize\footnotesize{7pt}{10}}
\makeatother

\renewcommand{\thefootnote}{\fnsymbol{footnote}}
\renewcommand\footnoterule{\vspace*{1pt}%
\color{cream}\hrule width 3.5in height 0.4pt \color{black}\vspace*{5pt}} 
\setcounter{secnumdepth}{5}

\makeatletter 
\renewcommand\@biblabel[1]{#1}            
\renewcommand\@makefntext[1]%
{\noindent\makebox[0pt][r]{\@thefnmark\,}#1}
\makeatother 
\renewcommand{\figurename}{\small{Fig.}~}
\sectionfont{\sffamily\Large}
\subsectionfont{\normalsize}
\subsubsectionfont{\bf}
\setstretch{1.125} 
\setlength{\skip\footins}{0.8cm}
\setlength{\footnotesep}{0.25cm}
\setlength{\jot}{10pt}
\titlespacing*{\section}{0pt}{4pt}{4pt}
\titlespacing*{\subsection}{0pt}{15pt}{1pt}

\fancyfoot{}
\fancyfoot[LO,RE]{\vspace{-7.1pt}\includegraphics[height=9pt]{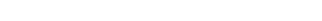}}
\fancyfoot[CO]{\vspace{-7.1pt}\hspace{13.2cm}\includegraphics{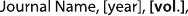}}
\fancyfoot[CE]{\vspace{-7.2pt}\hspace{-14.2cm}\includegraphics{head_foot/RF}}
\fancyfoot[RO]{\footnotesize{\sffamily{1--\pageref{LastPage} ~\textbar  \hspace{2pt}\thepage}}}
\fancyfoot[LE]{\footnotesize{\sffamily{\thepage~\textbar\hspace{3.45cm} 1--\pageref{LastPage}}}}
\fancyhead{}
\renewcommand{\headrulewidth}{0pt} 
\renewcommand{\footrulewidth}{0pt}
\setlength{\arrayrulewidth}{1pt}
\setlength{\columnsep}{6.5mm}
\setlength\bibsep{1pt}

\makeatletter 
\newlength{\figrulesep} 
\setlength{\figrulesep}{0.5\textfloatsep} 

\newcommand{\topfigrule}{\vspace*{-1pt}%
\noindent{\color{cream}\rule[-\figrulesep]{\columnwidth}{1.5pt}} }

\newcommand{\botfigrule}{\vspace*{-2pt}%
\noindent{\color{cream}\rule[\figrulesep]{\columnwidth}{1.5pt}} }

\newcommand{\dblfigrule}{\vspace*{-1pt}%
\noindent{\color{cream}\rule[-\figrulesep]{\textwidth}{1.5pt}} }

\makeatother

\twocolumn[
  \begin{@twocolumnfalse}
{\includegraphics[height=30pt]{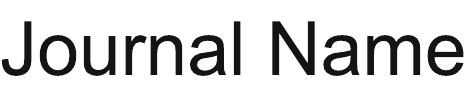}\hfill\raisebox{0pt}[0pt][0pt]{\includegraphics[height=55pt]{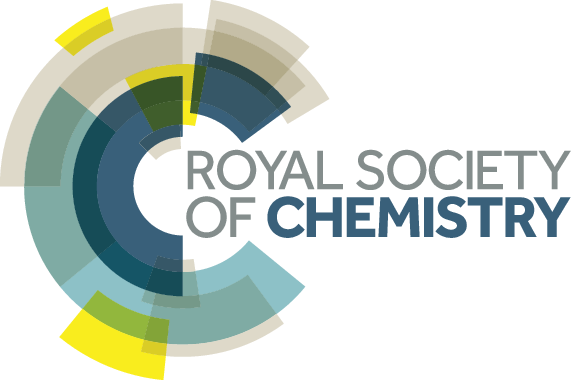}}\\[1ex]
\includegraphics[width=18.5cm]{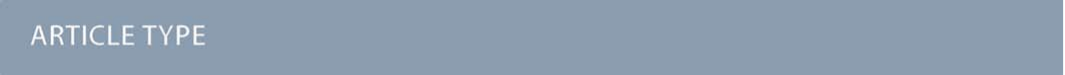}}\par
\vspace{1em}
\sffamily
\begin{tabular}{m{4.5cm} p{13.5cm} }

\includegraphics{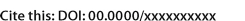} & \noindent\LARGE{\textbf{Diffusiophoresis in porous media saturated with a mixture of electrolytes}} \\
\vspace{0.3cm} & \vspace{0.3cm} \\

 & \noindent\large{Siddharth Sambamoorthy$^a$ and Henry C. W. Chu$^{b\ast}$} \\

\vspace{-0.5cm}

\includegraphics{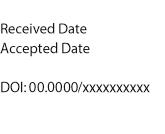} & \noindent\normalsize{Current theories of diffusiophoresis in porous media are limited to a porous medium saturated with a valence symmetric electrolyte.  A predictive model for diffusiophoresis in porous media saturated with a valence asymmetric electrolyte, or a general mixture of valence symmetric and asymmetric electrolytes, is lacking.  To close this knowledge gap, in this work we develop a mathematical model, based upon the regular perturbation method and numerical integration, to compute the diffusiophoretic mobility of a colloid in porous media saturated with a general mixture of electrolytes.  We model the electrokinetics using the Poisson-Nernst-Planck equations and the fluid transport in porous media using the Brinkman equation with an electric body force.  We report three novel key findings.  First, we demonstrate that, in the same electrolyte concentration gradient, lowering the permeability of the porous medium can significantly weaken the colloid diffusiophoretic motion.  Second, we show that, surprisingly, by using a valence asymmetric electrolyte the colloid diffusiophoretic motion in a denser porous medium can be stronger than that in a less dense porous medium saturated with a symmetric electrolyte.  Third, we demonstrate that varying the composition of an electrolyte mixture does not only change the strength of the colloid diffusiophoretic motion drastically, but also qualitatively its direction.  The model developed from this work can be used to understand and predict natural phenomena such as intracellular transport, as well as design technological applications such as enhanced oil recovery, nanoparticle drug delivery, and colloidal species separation.} 

\end{tabular}

 \end{@twocolumnfalse} \vspace{0.6cm}

  ]

\renewcommand*\rmdefault{bch}\normalfont\upshape
\rmfamily
\section*{}
\vspace{-1cm}

\footnotetext{\textit{$^{a}$~Department of Chemical Engineering, University of Florida, Gainesville, FL 32611, United States; $^{b}$~Department of Chemical Engineering and Department of Mechanical and Aerospace Engineering, University of Florida, Gainesville, FL 32611, United States; E-mail: h.chu@ufl.edu}}




\section{\label{sec:Sec1}Introduction}
Diffusiophoresis is the deterministic motion of a colloid induced by a surrounding solute concentration gradient.\citep{Prieve84, Velegol16, Keh16, Lee18Book, Shin20, Shim22}  Diffusiophoresis in an ionic solute concentration gradient comprises the chemiphoretic and electrophoretic transport: the chemiphoretic transport is driven by an osmotic pressure gradient and the electrophoretic transport is driven by an electric field; both the osmotic pressure gradient and the electric field are induced by the solute concentration gradient.  Diffusiophoresis is pivotal in natural settings such as intracellular transport\citep{Loose08, Lutkenhaus08, Sear19, Ramm21} and is increasingly utilized in applications such as enhanced oil recovery,\citep{Kar15, Park21, Shi21, Tan21} nanoparticle drug delivery,\citep{Shin19, Doan21} and colloidal species separation.\citep{Lee14, Shi16, Shin16, Zhai16, Shin17, Raynal18, Raynal19, Ault19, McMullen19, Ha19, Chu20a, Rasmussen20, Chu21, Singh20, Chu22, Alessio22, McKenzie22, Volk22, Xu23, Akdeniz23, Zhang24a, Teng23, Zhang24b}  The diffusiophoretic velocity follows $\boldsymbol{u} = M \nabla \log n$, where $n$ is the ion number density and $M$ is the diffusiophoretic mobility.  A positive (negative) $M$ corresponds to a colloid driven up or down the solute concentration gradient.  The mobility is the key to quantify diffusiophoresis and encompasses physical properties of the colloid and the surrounding media.  The objective of this article is to develop a predictive model of diffusiophoresis that accounts for the surrounding porous media and a mixture of valence symmetric and asymmetric electrolytes, thereby uncovering and predicting their impacts on diffusiophoresis.

Current theories of diffusiophoresis in a valence asymmetric electrolyte and a mixture of valence symmetric and asymmetric electrolytes have ignored the presence of porous media.  \citet{Pawar93} developed the first theory of diffusiophoresis in a valence asymmetric electrolyte, where the colloid electric double layer thickness $\kappa^{-1}$ is infinitesimally thin relative to the colloid radius $a$, $i.e.$, $\kappa a \gg 1$.  Bhattacharyya and co-workers\citep{Majhi22, Bhaskar23a} derived a general model that accounts for arbitrary values of the colloid surface potential $\zeta$ and $\kappa a$.  \citet{Wilson20} measured the colloid diffusiophoretic mobility in a valence asymmetric electrolyte in experiments and obtained good agreement with prior theories.  \citet{Chiang14} developed the first theory of diffusiophoresis in a mixture of electrolytes.  Their theory assumes that $\kappa a \gg 1$ and $\zeta$ is low compared to the thermal voltage $kT/e$, where $k$ is the Boltzmann constant, $T$ is the absolute temperature, and $e$ is the proton charge.  \citet{Shi16} derived a more general expression for the diffusiophoretic mobility, by relaxing the assumption of $\zeta \ll kT/e$.  \citet{Gupta19} and \citet{Ohshima21} presented approximate expressions for the mobility in the limit of $\zeta \ll 1$, $\zeta \gg 1$, and $\kappa a \geq 50$.  \citet{Samanta23} derived a general model that accounts for $\kappa a \geq 50$ and arbitrary values of $\zeta$.

Current theories of diffusiophoresis in porous media are not applicable to a valence asymmetric electrolyte nor a mixture of electrolytes, and are limited to a valence symmetric electrolyte, $i.e.$, two ionic species with valence $z_1 = -z_2 = z$.  Recently, our group pioneered a theory for diffusiophoresis in porous media \citep{Sambamoorthy23}.  The theory models the hydrodynamic interactions between the colloid and the porous media, which act to dampen the colloid motion as the porous medium permeability decreases.  The model accurately captures diffusiophoresis experiments by \citet{Doan21} with no adjustable parameters, where collagen gels are used as the porous media.  \citet{Bhaskar23b} incorporated the effect of ion size into the theory, showing quantitatively different predictions.  \citet{Somasundar23} demonstrated the feasibility of diffusiophoresis in another porous medium, by driving colloids through porous bacterial films using a concentration gradient of a valence symmetric electrolyte.  However, currently it is lacking a model that can simulate diffusiophoresis in porous media saturated with a valence asymmetric electrolyte or a general mixture of valence symmetric and asymmetric electrolytes.

In this work, we develop a mathematical model to predict the colloid diffusiophoretic mobility in a porous medium saturated with a general mixture of valence symmetric and asymmetric electrolytes.  We employ the Poisson-Nernst-Planck equations to model the ion transport and electric potential distribution, and the Brinkman equation with an electric body to model the fluid transport in a porous medium.  We solve the equations using a combination of the regular perturbation method and numerical integration.  We report three novel key findings that highlight the coupled effects of porous media and electrolyte mixtures on diffusiophoresis.  \textcolor{black}{First, we show that in the same electrolyte concentration gradient the colloid diffusiophoretic motion is significantly weaker in a less permeable porous medium.  This is consistent with the fundamental nature of porous media in dampening hydrodynamics,\citep{Brinkman49, Happel83} and generalizes the same conclusion that was made in prior work for a valence symmetric electrolyte\citep{Doan21, Sambamoorthy23, Bhaskar23b, Somasundar23} to a valence asymmetric electrolyte and a general mixture of electrolytes.}  \textcolor{black}{Second, we show that by using a valence asymmetric electrolyte, surprisingly, diffusiophoresis in a denser porous medium can be stronger than that in a less dense porous medium filled with a valence symmetric electrolyte.  This demonstrates the competition between electrokinetics and hydrodynamics, and offers new insights to generate strong diffusiophoresis in porous media using a valence asymmetric electrolyte.}  \textcolor{black}{Third, we show that, in a mixture of electrolytes, not just the magnitude of the colloid diffusiophoretic motion but qualitatively its direction can change by varying the mixture composition.  This highlights the novelty of the present work to leverage valence asymmetric electrolyte and electrolyte mixtures to generate a richer set of diffusiophoresis responses.}

The rest of this article is structured as follows.  In Section~\ref{sec:Sec2}, we present the problem formulation for the electrokinetic equations and the diffusiophoretic mobility.  In Section~\ref{sec:Sec3}, we present the results and discussion that elucidate the above-mentioned three key findings.  In Section~\ref{sec:Sec4}, we conclude this study.

\section{\label{sec:Sec2}Problem formulation}
\begin{figure}[!htb]
	\centering
{\includegraphics[width=0.45\linewidth]{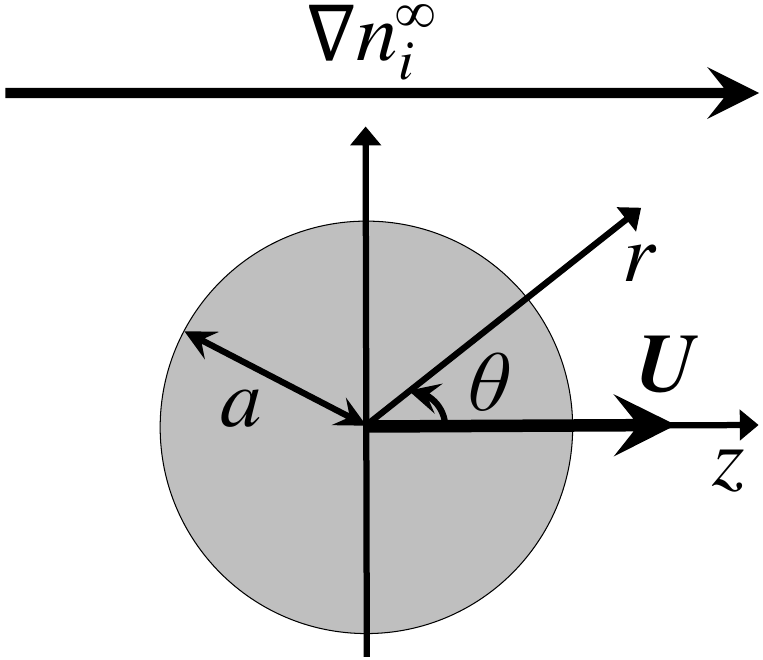}}
	\caption{A concentration gradient of a mixture of $N$ ionic species, $\nabla n_i^\infty$ with $i = 1, 2, ..., N$, induces the diffusiophoretic motion of a colloid of radius $a$ in a static porous medium.  The colloid diffusiophoretic velocity $\boldsymbol{U}$ is parallel to $\nabla n_i^\infty$.}
	\label{fig:Fig1}
\end{figure}

Consider a non-conducting colloid of radius $a$ and a constant surface charge density $q$ (or a constant surface potential $\zeta$) in a static porous medium with a constant permeability $l^2$ and screening length $l$, subject to a constant concentration gradient of an electrolyte mixture $\nabla n_i^\infty$ with $i = 1, 2, ..., N$ (Fig.~\ref{fig:Fig1}).  By symmetry, the colloid translates with a constant diffusiophoretic velocity $\boldsymbol{U}$ parallel to $\nabla n_i^\infty$ along the $z$-direction, where $\boldsymbol{U}$ is to be determined.  A reference frame moving with $\boldsymbol{U}$ is adopted.

At steady state, ionic species conservation requires that\citep{Hunter81}
\begin{equation}
    \label{eq:Eq2_1}
     \nabla \cdot \boldsymbol{j}_i = 0 \quad \text{with} \quad \boldsymbol{j}_i = -D_i \nabla n_i - \frac{D_i z_i e}{k T} n_i \nabla \phi + n_i \boldsymbol{u},
\end{equation}
where $\boldsymbol{j}_i$, $D_i$, $n_i$, and $z_i$ are the flux density, diffusivity, number density, and valence of the $i$-species, respectively; $\phi$ is the electric potential, and $\boldsymbol{u}$ is velocity of the electrolyte mixture.  The electric potential follows the Poisson equation\citep{Hunter81}
\begin{equation}
    \label{eq:Eq2_2}
     -\epsilon \nabla^2 \phi = \rho = \sum^N_{i = 1} z_i e n_i,
\end{equation}
where $\epsilon$ and $\rho$ are the permittivity and space charge density of the electrolyte mixture, respectively.  The fluid dynamics is governed by the continuity equation and the Brinkman equation with an electric body force\citep{Allison07, Tsai11, Hsu12, Hill16, Sambamoorthy23, Bhaskar23b}
\begin{equation}
    \label{eq:Eq2_3}
     \nabla \cdot \boldsymbol{u} = 0 \quad \text{and} \quad \boldsymbol{0} = -\nabla p + \eta\nabla^2\boldsymbol{u} - \rho \nabla\phi - \eta l^{-2} (\boldsymbol{u} + \boldsymbol{U}),
\end{equation}
where $p$ is the pressure and $\eta$ is the dynamic viscosity of the electrolyte mixture.

Eqn~\eqref{eq:Eq2_1}-\eqref{eq:Eq2_3} are specified by the following boundary conditions.  At the colloid surface at $r = a$, no slip and no penetration of the solvent prescribe $\boldsymbol{u} = \boldsymbol{0}$.  No penetration of the ionic species prescribes $\boldsymbol{n} \cdot \boldsymbol{j}_i = 0$, where $\boldsymbol{n}$ is the unit normal vector pointing away from the collois surface.  The colloid surface charge density $q$ or surface potential $\zeta$ could be specified as $-\boldsymbol{n} \cdot \epsilon \nabla \phi = q$ or $\phi = \zeta$.  Far from the colloid at $r \rightarrow \infty$, $\boldsymbol{u} \rightarrow -\boldsymbol{U}$ and $p \rightarrow p^\infty$, where $p^\infty$ is a reference constant pressure.  The ion number densities and their imposed gradients are represented by $n_i \rightarrow n_i^\infty + \nabla n_i^\infty \cdot \boldsymbol{r}$, where $n_i^\infty$ is the constant bulk electrolyte concentration and the position vector $\boldsymbol{r}$ is anchored at the centroid of the colloid.  To main bulk electroneutrality, the electrolyte concentration gradient induces an electric field $\boldsymbol{E} = -\nabla \phi = (kT/e)\beta \boldsymbol{G}$,\citep{Prieve84, Velegol16, Keh16} where $\boldsymbol{G} = (\nabla n_1^\infty)/n_1^\infty = ... = (\nabla n_N^\infty)/n_N^\infty$ and $\beta = (\sum^N_{i=1} z_i n_i^\infty D_i) / (\sum^N_{i=1} z_i^2 n_i^\infty D_i)$.  The ion diffusivity ratio $\beta$ controls the magnitude of $\boldsymbol{E}$ and thus the electrophoretic contribution in diffusiophoresis.

In typical diffusiophoresis,\citep{Prieve84, Velegol16, Keh16, Shin20, Shim22} the electrolyte concentration gradient at the size of the colloid is much smaller than the background concentration, $\alpha = |\boldsymbol{G}|a \ll 1$.  Using the regular perturbation method, we expand the dependent variables in $\alpha$ as
  \begin{equation}
  \label{eq:Eq2_4}
  \begin{gathered}
  n_i = n^\infty \left( \hat{n}^0_i + \alpha \hat{n}^1_i \right), \quad \phi = \frac{kT}{e} \left( \hat{\phi}^0 + \alpha \hat{\phi}^1 \right),\\
  \boldsymbol{u}=\frac{\epsilon k^2T^2}{e^{2}\eta a} \left( \hat{\boldsymbol{u}}^0 + \alpha \hat{\boldsymbol{u}}^1 \right), \quad p = \frac{\epsilon k^2T^2}{e^2a^2} \left( \hat{p}^0 + \alpha \hat{p}^1 \right),
\end{gathered}
\end{equation}
where $n^\infty = 2I = \sum^N_{i=1} z_i^2 n_i^\infty$ is twice the ionic strength of the mixture.  Quantities with carets are dimensionless.  We non-dimensionalize lengths by the colloid radius $a$ and the surface charge density by $\epsilon kT /(ea)$.  Substituting eqn~\eqref{eq:Eq2_4} into the governing equations and boundary conditions yield a set of differential equations at different orders of $\alpha$.  Below we present the $O(1)$ perturbation equations and the $O(\alpha)$ perturbation equations for computing the diffusiophoretic velocity and mobility.

\subsection{\label{sec:Sec2_1}$O(1)$ perturbation equations}
The $O(1)$ equations pertains to the equilibrium state where there is no electrolyte concentration gradient and fluid flow.  The ionic species conservation and Poisson equation are
\begin{equation}
    \label{eq:Eq2_5}
    \hat{\nabla} \cdot \left( \hat{\nabla} \hat{n}^0_i + z_i \hat{n}^0_i \hat{\nabla} \hat{\phi}^0 \right) = 0,
\end{equation}
\begin{equation}
    \label{eq:Eq2_6}
    \hat{\nabla}^2 \hat{\phi}^0 = -\hat{\kappa}^2 \sum_{i=1}^N z_i \hat{n}_i^0,
\end{equation}
where $\hat{\kappa} = \kappa a$ with $\kappa^{-1} = (\epsilon kT/e^2 n^\infty)^{1/2}$ being the Debye length.  The boundary condition at the colloid surface at $\hat{r} = 1$ is $-\boldsymbol{n} \cdot \hat{\nabla} \hat{\phi}^0 = \hat{q}$ or $\hat{\phi}^0 = \hat{\zeta}$.  The far-field boundary condition at $\hat{r} \rightarrow \infty$ are $\hat{n}_i^0 \rightarrow \hat{n}_i^\infty$ and $\hat{\phi}^0 \rightarrow 0$.  Integrating eqn~\eqref{eq:Eq2_5} with the boundary condition at $\hat{r} \rightarrow \infty$ gives the Boltzmann distribution of the ionic species $\hat{n}^0_i = \hat{n}^\infty_i \exp(-z_i \hat{\phi}^0)$.  Substituting this result into eqn~\eqref{eq:Eq2_6} gives the Poisson-Boltzmann equation\citep{Hunter81}
\begin{equation}
    \label{eq:Eq2_7}
   \frac{1}{\hat{r}^2} \frac{d}{d\hat{r}} \left(\hat{r}^2 \frac{d \hat{\phi}^0}{d\hat{r}}\right) = - \hat{\kappa}^2 \sum_{i=1}^N z_i \hat{n}^\infty_i \exp(-z_i \hat{\phi}^0).
\end{equation}

To solve eqn~\eqref{eq:Eq2_7} in a finite computational domain, we consider a sphere of radius $\hat{R}$ that is concentric to and encloses the colloid.\citep{OBrien78, OBrien81, Prieve87, Sambamoorthy23}  At a sufficiently large $\hat{R} = 1 + 20/\hat{\kappa}$ as in this work, the electric potential decays to zero asymptotically as $(1/\hat{r}) \exp(-\hat{\kappa} \hat{r})$.  The far-field boundary conditions that were at $\hat{r} \rightarrow \infty$ are transformed to $d \hat{\phi}^0 / d\hat{r} + (\hat{\kappa} + 1/\hat{R}) \hat{\phi}^0 = 0$ at $\hat{r} = \hat{R}$.  Eqn~\eqref{eq:Eq2_7} is solved using the standard shooting method with the boundary condition at $\hat{r} = 1$ and the transformed far-field boundary condition.

\subsection{\label{sec:Sec2_2}$O(\alpha)$ perturbation equations}
We convert the $O(\alpha)$ equations to a set of ordinary differential equations by introducing the $\hat{\psi}^{1}_i$ potential, $\hat{n}^1_i = -z_i \hat{n}^0_i (\hat{\psi}^1_i + \hat{\phi}^1)$,\citep{OBrien78, OBrien81, Sambamoorthy23} and writing the dependent variables as $\hat{u}^1_{\hat{r}}(\hat{r},\theta) = -(2/\hat{r}) \hat{h} \cos\theta$, $\hat{u}^1_\theta(\hat{r},\theta) = (1/\hat{r}) [d(\hat{r}\hat{h})/d\hat{r}] \sin\theta$, and $\hat{\psi}_i^1(\hat{r},\theta) = \hat{\Psi}_i^1 \cos\theta$, where $\hat{u}^1_{\hat{r}}$ and $\hat{u}^1_\theta$ are the radial and angular components of $\boldsymbol{\hat{u}}^1$, respectively.  The $O(\alpha)$ equations now reduce to solving for $\hat{h} = \hat{h}(\hat{r})$ and $\hat{\Psi}_i^1 = \hat{\Psi}_i^1(\hat{r})$, which are governed by
 \begin{equation}
    \label{eq:Eq2_8}
      \frac{d^2\hat{\Psi}^1_i}{d\hat{r}^2} + \frac{2}{\hat{r}}\frac{d\hat{\Psi}^1_i}{d\hat{r}} - \frac{2}{\hat{r}^2}\hat{\Psi}^1_i - z_i \frac{d\hat{\phi}^0}{d\hat{r}}\frac{d\hat{\Psi}^1_i}{d\hat{r}} + 2 Pe_i\frac{d\hat{\phi}^0}{d\hat{r}} \frac{\hat{h}}{\hat{r}} = 0,
\end{equation}
\begin{equation}
    \label{eq:Eq2_9}
    \begin{split}
     \frac{d^4 \hat{h}}{d\hat{r}^4} + & \frac{4}{\hat{r}}\frac{d^3 \hat{h}}{d\hat{r}^3} - \frac{4}{\hat{r}^2} \frac{d^2 \hat{h}}{d\hat{r}^2} - \gamma^2 \left( -\frac{2}{\hat{r}^2}\hat{h} + \frac{2}{\hat{r}} \frac{d\hat{h}}{d\hat{r}} + \frac{d^2 \hat{h}}{d\hat{r}^2} \right) +\\
     &\frac{\hat{\kappa}^2}{\hat{r}} \frac{d\hat{\phi}^0}{d\hat{r}} \sum_{i=1}^N z^2_i \hat{n}^0_i \hat{\Psi}^1_i = 0,
     \end{split}
\end{equation}
where $\gamma = a/l$ and the Peclet number $Pe_i \equiv \epsilon k^2 T^2 / (\eta e^2 D_i)$ describes the ratio of the advective to diffusive transport of the $i$-species.  The boundary conditions at the colloid surface at $\hat{r} = 1$ are $d\hat{\Psi}^1_i / d\hat{r} = 0$, $\hat{h} = 0$, and $d\hat{h} / d\hat{r} = 0$.  The far-field boundary conditions at $\hat{r} \rightarrow \infty$ are $\hat{\Psi}^1_i \rightarrow (\beta - 1/z_i) \hat{r}$ and $\hat{h} \rightarrow \hat{M} \hat{r}/2$.  The non-dimensionalized diffusiophoretic mobility $\hat{M}$ relates to the dimensional diffusiophoretic mobility $M$ via
\begin{equation}
    \label{eq:Eq2_10}
    \hat{M} = \frac{\hat{U}}{\alpha} = \frac{e^2 \eta}{\epsilon k^2T^2} M.
\end{equation}

\begin{figure*}[!htb]
	\centering
{\includegraphics[width=0.95\linewidth]{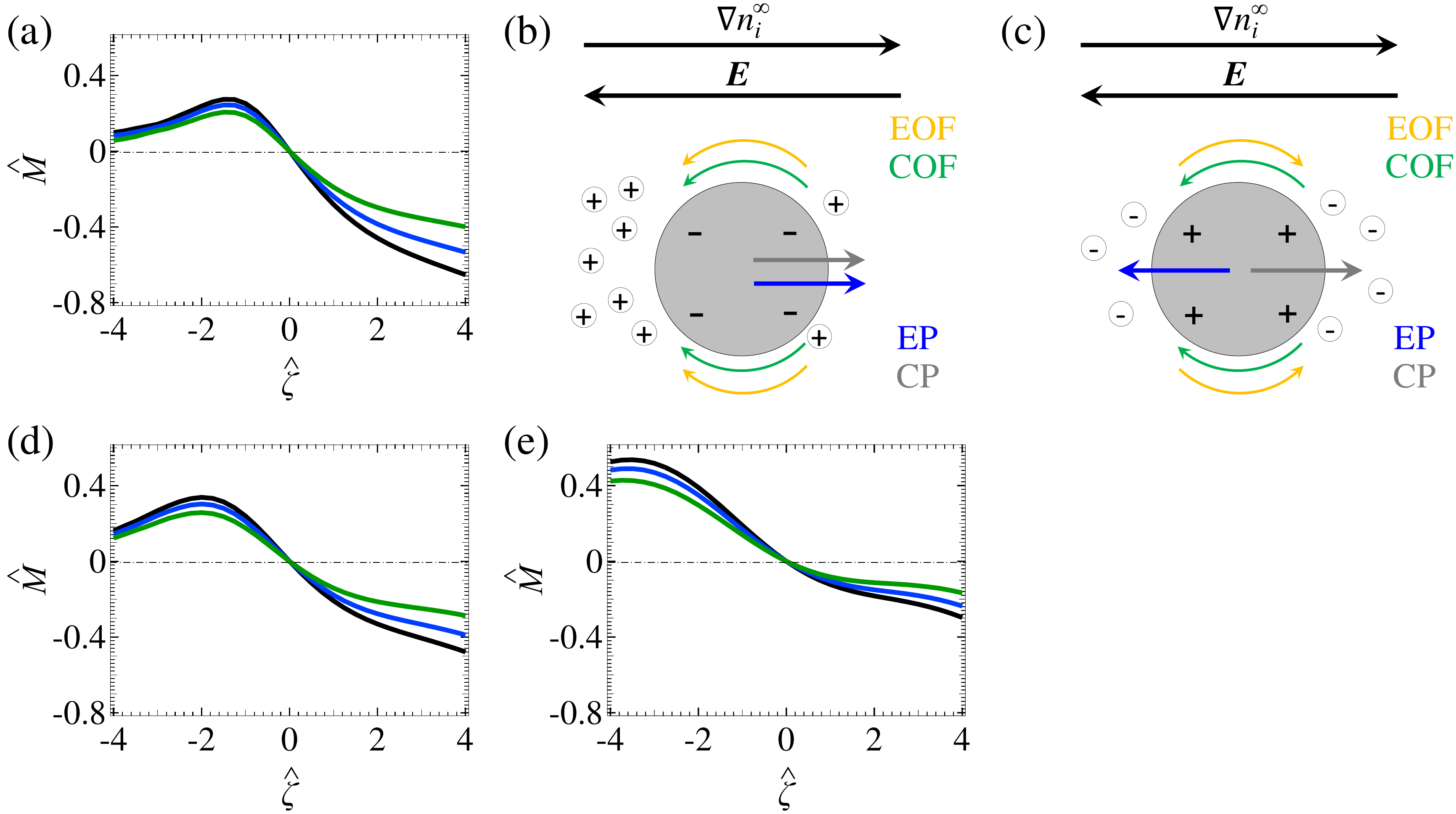}}
	\caption{(a) The non-dimensionalized  diffusiophoretic mobility $\hat{M}$ versus the non-dimensionalized colloid surface potential $\hat{\zeta}$ in different porous media saturated with a $\text{AlCl}_3$ solution of ionic strength $I = 0.25 \; \text{mM}$.  Black line: an electrolyte solution in the absence of porous media (the ratio of the screening length to the colloid radius $l/a \gg 1$); Blue line: a less dense porous medium ($l/a = 1$); Green line: a denser porous medium ($l/a = 0.5$).  (b) A negatively charged colloid undergoing diffusiophoresis moving in the positive $z$-direction (to the right).  The induced electric field $\boldsymbol{E}$ points to the negative $z$-direction.  Both the electroosmotic flow (EOF) and chemiosmotic flow (COF) transport ions in the negative $z$-direction.  Both electrophoresis (EP) and chemiphoresis (CP) in diffusiophoresis move the colloid in the positive $z$-direction.  (c) A positively charged colloid undergoing diffusiophoresis moving in the negative $z$-direction (to the left).  The EOF transports ions in the positive $z$-direction and the COF transports ions in the negative $z$-direction.  CP is outweighed by EP which acts in the negative $z$-direction.  (d) $\hat{M}$ versus $\hat{\zeta}$ in different porous media saturated with a $\text{BaCl}_2$ solution.  (e) $\hat{M}$ versus $\hat{\zeta}$ in different porous media saturated with a $\text{NaCl}$ solution.  The color scheme and the ionic strength of (d) and (e) are the same as (a).}
	\label{fig:Fig2}
\end{figure*}

To solve eqn~\eqref{eq:Eq2_8} and \eqref{eq:Eq2_9} in a finite computational domain, it again requires transforming the far-field boundary conditions at $\hat{r} \rightarrow \infty$ to ones at $\hat{r} = \hat{R}$.  First, in the asymptotic limit of a vanishing electric potential at a sufficiently large $\hat{R}$, the far-field boundary condition for $\hat{\Psi}^1_i$ that was at $\hat{r} \rightarrow \infty$ is transformed to be
\begin{equation}
    \label{eq:Eq2_11}
    2\hat{\Psi}^1_i + \hat{R} \frac{d\hat{\Psi}^1_i}{d\hat{r}} = 3 \left(\beta - \frac{1}{z_i} \right)\hat{R} \quad \text{at} \quad \hat{r} = \hat{R}.
\end{equation}
Second, the net hydrodynamic and electric forces are zero on the large sphere of radius $\hat{R}$ that encloses a freely suspended colloid.  A sufficiently large $\hat{R}$ guarantees that the sphere is electric force-free.\citep{OBrien78, OBrien81, Prieve87, Sambamoorthy23}  Thus, the constraint reduces to that the sphere is hydrodynamic force-free: $\boldsymbol{\hat{F}} = 2\pi \hat{R}^2 \int_0^\pi \boldsymbol{\hat{\sigma}} \cdot \boldsymbol{n} \; \sin\theta d\theta - 2\pi\gamma^2 \int_{0}^{\pi} \int_1^{\hat{R}} \left(\boldsymbol{\hat{u}} + \boldsymbol{\hat{U}}\right)\hat{r}^2\sin\theta \; d\hat{r}\;d\theta  = \boldsymbol{0}$, where $\boldsymbol{\hat{\sigma}} = -\hat{p}\boldsymbol{I} +  \hat{\nabla} \boldsymbol{\hat{u}} + (\hat{\nabla} \boldsymbol{\hat{u}})^{T}$ is the Newtonian stress tensor and $\boldsymbol{I}$ is the identity tensor.  Rearranging eqn~\eqref{eq:Eq2_9} at $\hat{r} = \hat{R}$ with the hydrodynamic force-free condition yields
\begin{equation}
    \label{eq:Eq2_12}
        \begin{split}
   & \lambda \hat{R} \frac{d^2 \hat{h}}{d\hat{r}^2} - \left(4 \gamma^3 +3 \gamma^3 \hat{R}^3 + 9\gamma^2\hat{R}^2 +18 \gamma  \hat{R} + 18\right)\hat{h} + \\
   &\left[-6 \gamma^2 + 3 \gamma^3 \hat{R}^4 +9 \gamma^2 \hat{R}^3 + 18 \gamma \hat{R}^2 - 2 \left(\gamma^3 - 9\right) \hat{R}\right]\frac{d\hat{h}}{d\hat{r}} = 0, 
        \end{split}
\end{equation}
\begin{equation}
\label{eq:Eq2_13}
    \begin{split}
   &- 9 \gamma  \left(\gamma^2 \hat{R}^2+2 \gamma  \hat{R}+2\right)\hat{h}+
   3 \gamma  \left(-2 \gamma^2+3 \gamma^2 \hat{R}^3+ \right. \\
   &\left. 6 \gamma  \hat{R}^2+6 \hat{R} \right) \frac{d\hat{h}}{d\hat{r}} + \lambda \left[\hat{R}\frac{d^3 \hat{h}}{d \hat{r}^2} + (\gamma  \hat{R}+4) \frac{d^2 \hat{h}}{d\hat{r}^2}\right] = 0,
        \end{split}
\end{equation}
\begin{equation}
    \label{eq:Eq2_14}
    \hat{M} = \left[\frac{6 \left(\gamma^2 \hat{R}^2+2 \gamma  \hat{R}+2\right)}{\lambda}\right]\hat{h} + \left[\frac{6 \hat{R} (\gamma  \hat{R}+1)}{\lambda}\right]\frac{d\hat{h}}{d\hat{r}},
\end{equation}
where $\lambda = -2 \gamma^2+3 \gamma^2 \hat{R}^3 + 9 \gamma \hat{R}^2 + 9\hat{R}$.  In sum, we solve eqn~\eqref{eq:Eq2_8} and \eqref{eq:Eq2_9} for $\hat{h}$ and $\hat{\Psi}_i^1$ subject to the boundary conditions $d\hat{\Psi}^1_i / d\hat{r} = 0$, $\hat{h} = 0$, and $d\hat{h} / d\hat{r} = 0$ at $\hat{r} = 1$, in addition to the boundary conditions \eqref{eq:Eq2_11} to \eqref{eq:Eq2_13} at $\hat{r} = \hat{R}$, using the solver \textit{NDSolve} in Wolfram Mathematica.  We obtain the diffusiophoretic mobility from eqn~\eqref{eq:Eq2_14}.  As validation, these equations can recover results for (i) diffusiophoresis in porous media saturated with a valence symmetric electrolyte when $N = 2$ and $z_1 = -z_2 = z$,\citep{Sambamoorthy23, Bhaskar23b} (ii) diffusiophoresis in a mixture of valence symmetric and asymmetric electrolytes in the absence of porous media when $\gamma \rightarrow 0$,\citep{Chiang14, Shi16, Gupta19, Ohshima21, Samanta23} (iii) diffusiophoresis in a valence asymmetric electrolyte when $N = 2$ and $z_1 \neq z_2$,\citep{Pawar93, Majhi22, Bhaskar23a} and (iv) diffusiophoresis in a valence symmetric electrolyte when $N = 2$ and $z_1 = -z_2 = z$.\citep{Prieve84, Prieve87, Keh16}  Some of these validations are presented in Section~\ref{sec:Sec3} to contrast with the new results of this work on diffusiophoresis in porous media saturated with a valence asymmetric electrolyte, and a mixture of valence symmetric and asymmetric electrolytes.

\section{\label{sec:Sec3}Results and discussion}
Current knowledge of diffusiophoresis in porous media is limited to porous media saturated with a valence symmetric electrolyte.\citep{Doan21, Sambamoorthy23, Bhaskar23b, Somasundar23}  In this section, we compute and discuss novel results diffusiophoresis in porous media saturated with a valence asymmetric electrolyte, in addition to a mixture of valence symmetric and asymmetric electrolytes.  Model inputs are taken from typical experiments:\citep{Hunter81, Righetti96, Pluen99, Kar15, Shin16, Doan21, Somasundar23} $T = 298 \;\text{K}$, $l \geq 50 \; \text{nm}$, $a = 100 \; \text{nm}$, $I \geq 10^{-2} \;\text{mM}$, and $\zeta \in [-100, 100] \; \text{mV}$ that corresponds to $\hat{\zeta} \in [-4, 4]$.  The porous medium pore diameter is at least about $1.5$ times the colloid diameter, which allows the passage of the colloid.\citep{Pluen99, Doan21}  In Section~\ref{sec:Sec3_1}, we present the diffusiophoretic mobilities $\hat{M}$ in porous media saturated with a valence asymmetric electrolyte.  \textcolor{black}{This will demonstrate the first and the second key findings of this work.  Namely, in the same electrolyte lowering the permeability of the porous medium significantly weakens the colloid diffusiophoretic motion; surprisingly, in an valence asymmetric electrolyte diffusiophoresis in a denser porous medium can be stronger than that in a less dense porous medium saturated with a valence symmetric electrolyte.}  In Section~\ref{sec:Sec3_2}, we present $\hat{M}$ in porous media saturated with a mixture of valence symmetric and asymmetric electrolytes.  \textcolor{black}{This will demonstrate the third key finding of this work.  Namely, in a mixture of electrolytes not just the magnitude but also the direction of the colloid motion can change by varying the mixture composition.}

\subsection{\label{sec:Sec3_1}Diffusiophoresis in porous media saturated with a valence asymmetric electrolyte}
Let us first examine diffusiophoresis in porous media saturated with a $\text{AlCl}_3$ solution.  The diffusivity of $\text{Al}^{3+}$ and $\text{Cl}^{-}$ are $D_{\text{Al}^{3+}} = 0.541 \times 10^{-9} \;\text{m}^2 \;\text{s}^{-1}$ and $D_{\text{Cl}^{-}} = 2.032 \times 10^{-9} \;\text{m}^2 \;\text{s}^{-1}$, respectively,\citep{Dane02} giving the ion diffusivity ratio $\beta = -0.408$.  Fig.~\ref{fig:Fig2}(a) shows the non-dimensionalized diffusiophoretic mobility $\hat{M}$ versus the non-dimensionalized colloid surface potential $\hat{\zeta}$.  The ionic strength of the solution is $I = 0.25 \;\text{mM}$.  Let us first focus on the black line with $l/a \gg 1$, which corresponds to an infinitely permeable porous media, $i.e.$, an electrolyte solution in the absence of porous media.  \textcolor{black}{Here, the first observation is that for a negatively charged colloid $\hat{M}$ is positive, whereas for a positively charged colloid $\hat{M}$ is negative.}  This is consistent with the classical understanding of diffusiophoresis as follows.\citep{Prieve84, Velegol16, Keh16, Lee18Book}  Let us first consider a negatively charged colloid as shown in Fig.~\ref{fig:Fig2}(b).  A negative $\beta$ implies that the induced electric field $\boldsymbol{E}$ is pointing in the negative $z$-direction.  The field $\boldsymbol{E}$ drives positive counterions to the left and generates an electroosmotic flow in the same direction, causing the colloid to undergo electrophoresis to the right.  Meanwhile, the osmotic pressure gradient induced by the electrolyte concentration gradient generates a chemiosmotic flow to the left, causing the colloid to undergo chemiphoresis to the right.  Since diffusiophoresis is the sum of electrophoresis and chemiphoresis, diffusiophoresis acts in the positive $z$-direction and $\hat{M}$ is positive.  In contrast, for a positively charged colloid as shown in Fig.~\ref{fig:Fig2}(c), electrophoresis which acts to the left outweighs chemiphoresis which acts to the right.  This results in diffusiophoresis acting in the negative $z$-direction and $\hat{M}$ is negative.

\textcolor{black}{The second observation of the black line in Fig.~\ref{fig:Fig2}(a) is that for a negatively charged colloid the magnitude of $\hat{M}$ experiences a decay at $\hat{\zeta} \leq 1$, whereas for a positively charged colloid the magnitude of $\hat{M}$ is monotonically increasing.}  This can be understood by examining the ionic transport between the two cases.  Let us first consider a negatively charged colloid by referring to Fig.~\ref{fig:Fig2}(b).  Both the electroosmotic and chemiosmotic flow are transporting positive counterions downstream to the colloid to the left.  Due to Coulombic attraction, these counterions will slow down the colloid motion that is to the right.\citep{OBrien78, Lee18Book}  The stronger the colloid's charge, the stronger this slowing-down effect to the colloid.  Therefore, $\hat{M}$ decays as $\hat{\zeta}$ becomes more negative.  Next, let us consider a positively charged colloid by referring to Fig.~\ref{fig:Fig2}(c).  In contrast to Fig.~\ref{fig:Fig2}(b), in Fig.~\ref{fig:Fig2}(c) only the electroosmotic flow is transporting negative counterions downstream to the colloid to the right, whereas the chemiosmotic flow is transporting counterions upstream to the colloid to the left.  As a result, the slowing-down effect to the colloid due to Coulombic attraction between the counterions and the colloid is weak.  Therefore, the magnitude of $\hat{M}$ does not experience a decay but grows monotonically.

\textcolor{black}{The first key finding of this work arises from an overview of Fig.~\ref{fig:Fig2}(a), where decreasing $l/a$ decreases the magnitude of $\hat{M}$ significantly.}  For example, at $\hat{\zeta} = 2$, decreasing $l/a$ (from black to green line) decreases the magnitude of $\hat{M}$ from $0.459$ to $0.298$.  Physically, decreasing $l/a$ implies decreasing the permeability of the porous medium to the transport of the electrolyte and the colloid.  As the permeability decreases, the hydrodynamic drag to the electrolyte and the colloid increases, and therefore $\hat{M}$ decreases.  This key finding is consistent with the fundamental nature of porous media in weakening hydrodynamics.\citep{Brinkman49, Happel83}  Also, this key finding and the two observations in the previous paragraphs hold in a lower valence, valence asymmetric electrolyte $\text{BaCl}_2$ as shown in Fig.~\ref{fig:Fig2}(d) ($D_{\text{Ba}^{2+}} = 0.847 \times 10^{-9} \;\text{m}^2 \;\text{s}^{-1}$\citep{Dane02} and $\beta = -0.318$) and in a valence symmetric electrolyte $\text{NaCl}$ as shown in Fig.~\ref{fig:Fig2}(e) ($D_{\text{Na}^{+}} = 1.33 \times 10^{-9} \;\text{m}^2 \;\text{s}^{-1}$\citep{Dane02} and $\beta = -0.207$).  As validation, the results with $l/a \gg 1$ in Fig.~\ref{fig:Fig2}(a), (d), and (e) can recover those in prior work for diffusiophoresis in the absence of porous media,\citep{Pawar93, Majhi22, Bhaskar23a} and the results with a finite $l/a$ in Fig.~\ref{fig:Fig2}(e) can recover those in prior work for diffusiophoresis in porous media.\citep{Sambamoorthy23, Bhaskar23b}

\begin{figure}[!htb]
	\centering
{\includegraphics[width=0.8\linewidth]{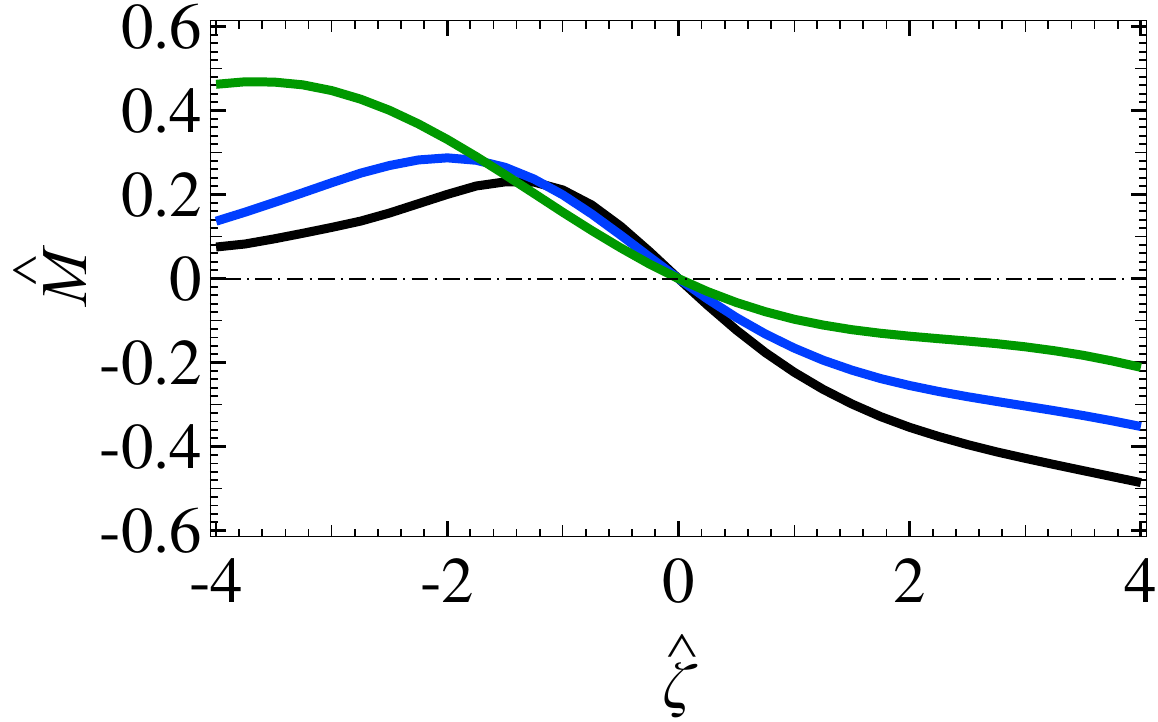}}
	\caption{The non-dimensionalized diffusiophoretic mobility $\hat{M}$ versus the non-dimensionalized colloid surface potential $\hat{\zeta}$ in the same porous media (the ratio of the screening length to the colloid radius $l/a = 0.75$) saturated with different electrolyte solutions of ionic strength $I = 0.25 \; \text{mM}$.  Black line: $\text{AlCl}_3$; Blue line: $\text{BaCl}_2$; Green line: $\text{NaCl}$.}
	\label{fig:Fig3}
\end{figure}

Next, let us turn to Fig.~\ref{fig:Fig3}.  We consider a porous medium with $l = 75 \;\text{nm}$.  In three difference cases, the porous medium is saturated with three different electrolyte solutions, $\text{AlCl}_3$, $\text{BaCl}_2$, and $\text{NaCl}$.  The ionic strength of the three solutions are the same as $I = 0.25 \; \text{mM}$.  \textcolor{black}{The third observation is that, for a negatively charged colloid, the decay in the diffusiophoretic mobility $\hat{M}$ occurs at a lower colloid surface potential $\hat{\zeta}$ in a higher valence electrolyte than in a lower valence electrolyte.}  For example, the decay in $\hat{M}$ for $\text{AlCl}_3$ (black line) occurs at $\hat{\zeta} \leq -1$, whereas the decay for $\text{NaCl}$ (green line) occurs at $\hat{\zeta} \leq -3.5$.  This can be understood by recalling the explanation for the second observation.  Specifically, the decay is due to the Coulombic attraction between the colloid and the downstream counterions.  For $\text{AlCl}_3$, the counterions are $\text{Al}^{3+}$, which have a higher charge density than the counterions $\text{Na}^{+}$ of $\text{NaCl}$.  Hence, at the same $\hat{\zeta}$, the Coulombic attraction between the colloid and $\text{Al}^{3+}$ is stronger than that between the colloid and $\text{Na}^{+}$.  As a result, the decay in $\hat{M}$ occurs at a weaker $\hat{\zeta}$ for $\text{AlCl}_3$.

\textcolor{black}{The fourth observation is from the range of a positive $\hat{\zeta}$ in Fig.~\ref{fig:Fig3}, where $\hat{M}$ can be increasingly enhanced by a more valence asymmetric electrolyte.}  The enhancement in $\hat{M}$ is significant.  For example, at $\hat{\zeta} = 2$, $\hat{M}$ for $\text{NaCl}$ is $-0.137$ (green line).  In a $\text{BaCl}_2$ solution, $\hat{M}$ has a $86\%$ increase and equals $-0.254$ (blue line).  In a $\text{AlCl}_3$ solution, $\hat{M}$ has a $158\%$ increase and equals $-0.354$ (black line).  As noted in the first observation, the electrophoretic component of diffusiophoresis dominates for a positively charged colloid.  The driving force for electrophoresis is the induced electric field $\boldsymbol{E} = (kT/e)\beta \boldsymbol{G}$, which is proportional to the ion diffusivity ratio $\beta$.  Thus, as the magnitude of $\beta$ increases from $|\beta| = 0.207$ ($\text{NaCl}$) to $|\beta| = 0.318$ ($\text{BaCl}_2$) and $|\beta| = 0.408$ ($\text{AlCl}_3$), the magnitude of $\hat{M}$ increases.

\begin{figure}[t]
	\centering
{\includegraphics[width=0.8\linewidth]{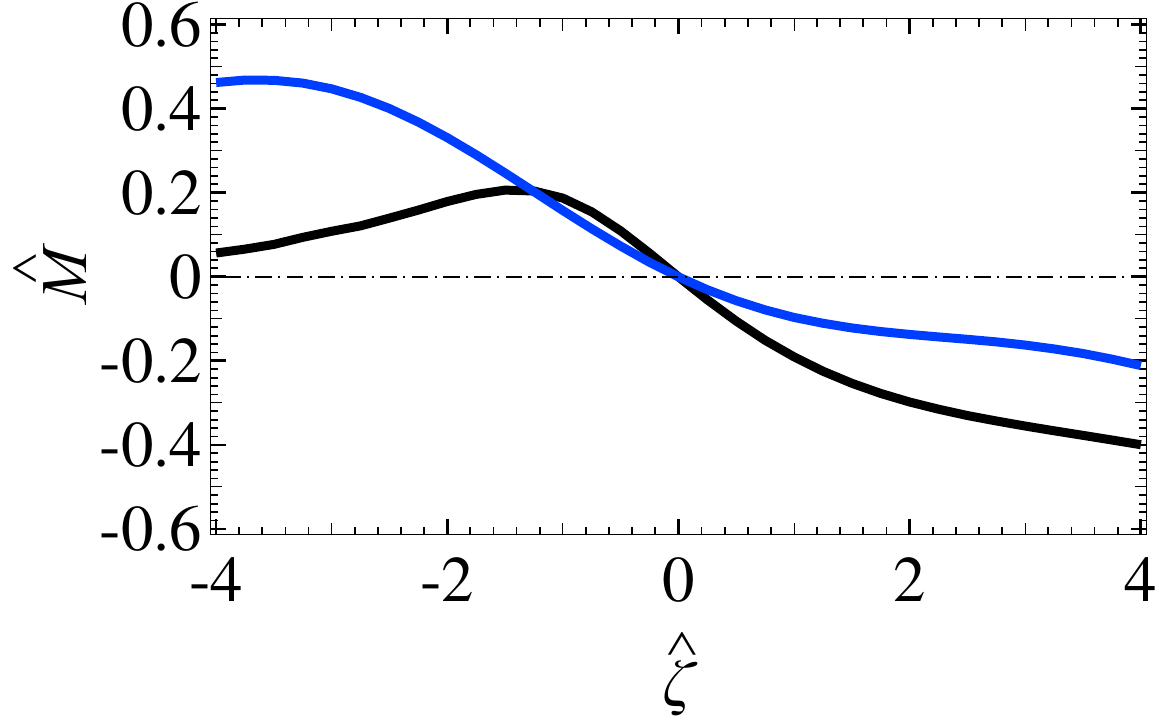}}
	\caption{The non-dimensionalized diffusiophoretic mobility $\hat{M}$ versus the non-dimensionalized colloid surface potential $\hat{\zeta}$ in different porous media saturated with different electrolyte solutions of ionic strength $I = 0.25 \; \text{mM}$.  Black line: a denser porous medium (the ratio of the screening length to the colloid radius $l/a = 0.5$) saturated with a $\text{AlCl}_3$ solution; Blue line: a less dense porous medium ($l/a = 0.75$) saturated with a $\text{NaCl}$ solution.}
	\label{fig:Fig4}
\end{figure}

\begin{figure*}[t]
	\centering
{\includegraphics[width=1\linewidth]{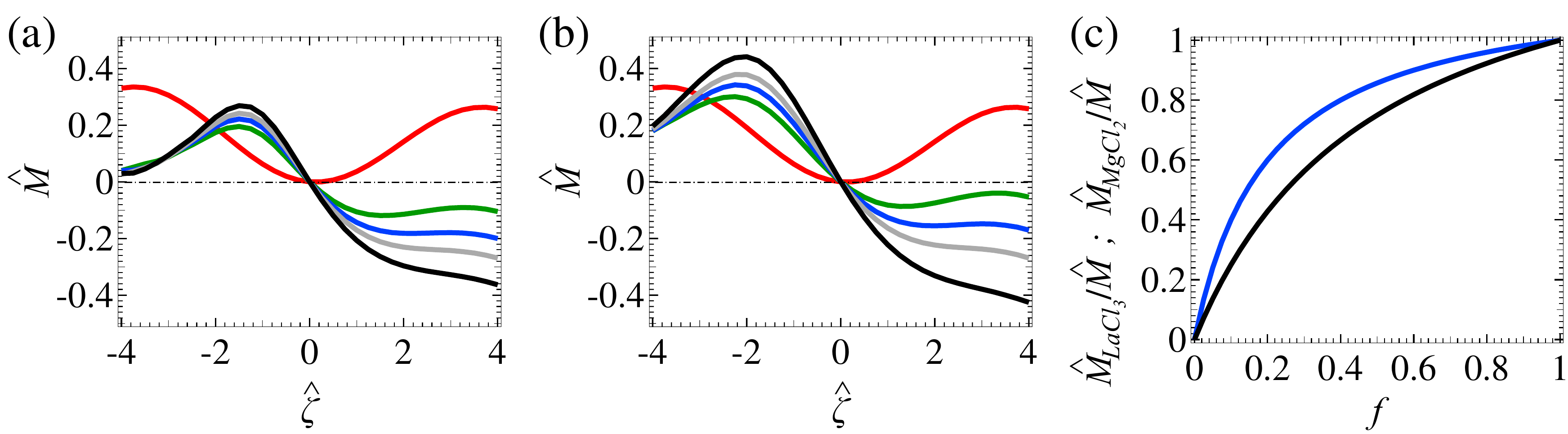}}
	\caption{(a) The non-dimensionalized diffusiophoretic mobility $\hat{M}$ versus the non-dimensionalized colloid surface potential $\hat{\zeta}$ in the same porous media (the ratio of the screening length to the colloid radius $l/a = 0.75$) saturated with solution mixtures of $p$ molar of $\text{KCl}$ and $q$ molar of $\text{LaCl}_3$ with different compositions $f = q / (p + q)$.  The solution ionic strength $I = 1 \; \text{mM}$.  Red line: $f = 0$; Green line: $f = 0.33$; Blue line: $f = 0.5$; Grey line: $f = 0.67$; Black line: $f = 1$.  (b) $\hat{M}$ versus $\hat{\zeta}$ in the same porous media ($l/a = 0.75$) saturated with solution mixtures of $p$ molar of $\text{KCl}$ and $q$ molar of $\text{MgCl}_2$ with different compositions $f = q / (p + q)$.  The color scheme and the ionic strength are the same as (a).  (c) The ratio of the contribution to the mobility by the asymmetric electrolyte to that by the mixture $\hat{M}_{\text{LaCl}_3}/\hat{M}$ and $\hat{M}_{\text{MgCl}_2}/\hat{M}$ versus $f$.  Blue line: $\hat{M}_{\text{LaCl}_3}/\hat{M}$; Black line: $\hat{M}_{\text{MgCl}_2}/\hat{M}$.}
	\label{fig:Fig5}
\end{figure*}

\begin{figure*}[t]
	\centering
{\includegraphics[width=1\linewidth]{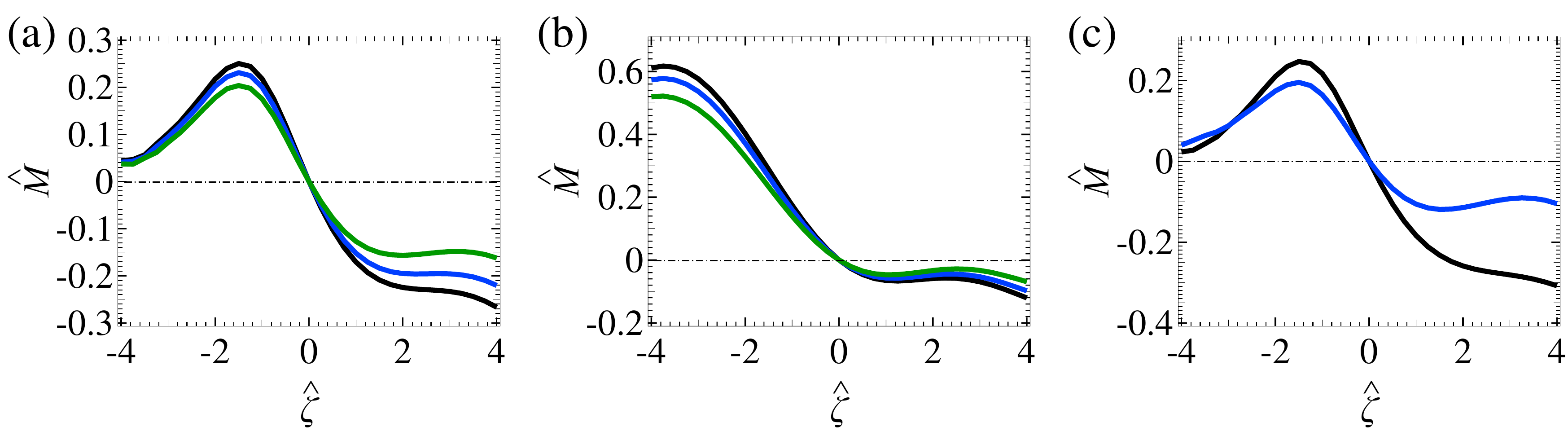}}
	\caption{(a) The non-dimensionalized diffusiophoretic mobility $\hat{M}$ versus the non-dimensionalized colloid surface potential $\hat{\zeta}$ in different porous media saturated with a solution mixture of $p$ molar of $\text{KCl}$ and $q$ molar of $\text{LaCl}_3$ with a composition $f = q / (p + q) = 0.5$.  The solution ionic strength $I = 1 \; \text{mM}$.  Black line: an electrolyte solution in the absence of porous media (the ratio of the screening length to the colloid radius $l/a \gg 1$); Blue line: a less dense porous medium $l/a = 1$; Green line: a denser porous medium $l/a = 0.5$.  (b) $\hat{M}$ versus $\hat{\zeta}$ in different porous media saturated with a phosphate-buffered saline (PBS) solution.  The color scheme and the ionic strength are the same as (a).  (c) $\hat{M}$ versus $\hat{\zeta}$ in different porous media saturated with different $\text{KCl}$/$\text{LaCl}_3$ mixtures.  Black line: a denser porous medium saturated with a higher fraction of valence asymmetric electrolyte $l/a = 0.5$ and $f =1$; Blue line: a less dense porous medium saturated with a lower fraction of valence asymmetric electrolyte $l/a = 0.75$ and $f =0.33$.}
	\label{fig:Fig6}
\end{figure*}

Next, let us turn to Fig.~\ref{fig:Fig4}.  The black line denotes a denser porous medium ($l/a = 0.5$) saturated with a $\text{AlCl}_3$ solution and the blue line denotes a less dense porous medium ($l/a = 0.75$) saturated with a $\text{NaCl}$ solution.  The solution ionic strength $I = 0.25 \; \text{mM}$ is the same in the two cases.  \textcolor{black}{Here, the second key finding of this work is that the magnitude of the diffusiophoretic mobility $\hat{M}$ in a denser porous medium can be stronger than that in a less dense porous medium.}  For example, for a positive colloid surface potential $\hat{\zeta}$, the magnitude of the black line is larger than that of the blue line.  This is a rather surprising and non-intuitive result, as one may expect that the colloid movement in a denser porous medium is always weaker than that in a less dense porous medium due to the stronger hydrodynamic drag to the fluid and the colloid.  However, results here show that, by employing a valence asymmetric electrolyte, diffusiophoresis in a denser porous medium can be much stronger than that in a less dense porous medium saturated with a valence symmetric electrolyte.  Physically, this is because the enhancement to the electrophoretic component of diffusiophoresis by the valence asymmetric electrolyte outweighs the hydrodynamic drag due to the porous media.  This finding offers new insights into using valence asymmetric electrolyte to generate strong diffusiophoresis in porous media.

\subsection{\label{sec:Sec3_2}Diffusiophoresis in porous media saturated with a mixture of valence symmetric and asymmetric electrolytes}
Let us examine diffusiophoresis in porous media saturated with solution mixtures of $p$ molar of $\text{KCl}$ and $q$ molar of $\text{LaCl}_3$ with different compositions $f = q / (p + q)$, where $f \in [0,1]$ with $f=0$ and $f=1$ representing a pure $\text{KCl}$ solution and a pure $\text{LaCl}_3$ solution, respectively.  The diffusivity of $\text{K}^+$ is $D_{\text{K}^+} = 1.96 \times 10^{-9} \;\text{m}^2 \;\text{s}^{-1}$ and the diffusivity of $\text{La}^{3+}$ is $D_{\text{La}^{3+}} = 0.619 \times 10^{-9} \;\text{m}^2 \;\text{s}^{-1}$.\citep{Dane02}  Fig.~\ref{fig:Fig5}(a) shows the diffusiophoretic mobility $\hat{M}$ versus the colloid surface potential $\hat{\zeta}$.  The screening length is $l = 75 \;\text{nm}$ and the ionic strength of the solution is $I = 1 \;\text{mM}$.  Let us first focus on the black and the red line with $f = 1$ and $f = 0$, respectively.  \textcolor{black}{The first observation is that the variation of $\hat{M}$ versus $\hat{\zeta}$ in a pure $\text{LaCl}_3$ solution ($f = 1$; black line) is qualitatively different from that in a pure $\text{KCl}$ solution ($f = 0$; red line).}  Specifically, in a pure $\text{LaCl}_3$ solution, $\hat{M}$ is positive for a negative $\hat{\zeta}$ and is negative for a positive $\hat{\zeta}$.  This follows the same trend and explanation as for $\text{AlCl}_3$ in the first observation in Section~\ref{sec:Sec3_1} and is not repeated here.  In contrast, in a pure $\text{KCl}$ solution, $\hat{M}$ is positive regardless of the sign of $\hat{\zeta}$.  This is understood by noting the very small value $\beta = -0.0188$, which reflects the negligibly small electrophoretic contribution in diffusiophoresis.  Hence, the colloid undergoes diffusiophoresis chiefly due to chemiphoresis that drives the colloid in the positive $z$-direction and thus $\hat{M}$ is positive.

\textcolor{black}{The third key finding of this work is that varying the mixture composition $f$ can change not just the magnitude of the colloid diffusiophoretic motion drastically, but also qualitatively its direction; the latter is reflected from the sign change of $\hat{M}$ for a fixed positive $\hat{\zeta}$ as $f$ increases.}  In a mixture of electrolytes, $\hat{M}$ is not simply the superposition nor the weighted average of the contributions from its individual electrolyte.  For example, in Fig.~\ref{fig:Fig5}(a) at $\hat{\zeta} = 3$, $\hat{M}_{f=0} = 0.240$ (red line) and $\hat{M}_{f=1} = -0.328$ (black line), but $\hat{M}_{f=0.5} = -0.179$ (blue line) $\neq (\hat{M}_{f=0} + \hat{M}_{f=1})/2$.  In fact, $\hat{M}$ depends nonlinearly on $f$.  This is because $\hat{M}$ is a result of the nonlinearly coupled Poisson-Nernst-Planck and Brinkman equations \eqref{eq:Eq2_1}-\eqref{eq:Eq2_3}.  Because of this intricate coupling, $\hat{M}$ can vary non-monotonically with $f$.  For example, at $\hat{\zeta} = -2.8$, $\hat{M}_{f=0} = 0.283$ (red line); as $f$ increases, $\hat{M}$ first decreases and then increases.  In the special case at $\hat{\zeta} = -2.4$, $\hat{M}_{f=0}$ (red line) can even be identical to $\hat{M}_{f=1}$ (black line), meaning that a colloid undergoes the same diffusiophoretic motion regardless of the electrolyte identity.  These observations are not restricted to a $\text{KCl}/\text{LaCl}_3$ mixture but is general to other electrolyte mixtures.  For example, Fig.~\ref{fig:Fig5}(b) shows a similar set of diffusiophoretic response in a $\text{KCl}/\text{MgCl}_2$ mixture ($D_{\text{Mg}^{2+}} = 0.706 \times 10^{-9} \;\text{m}^2 \;\text{s}^{-1}$\citep{Dane02}) as in Fig.~\ref{fig:Fig5}(a).  These observations underline the importance of this work to solve the coupled Poisson-Nernst-Planck-Brinkman equations to predict diffusiophoresis.

A scaling analysis illustrates that the nonlinearity between $\hat{M}$ and $f$ depends on the ion valence $z$.  First, referring to the far-field boundary condition below eqn~\eqref{eq:Eq2_9}, $\hat{M} \sim \hat{h}$.  Further, from eqn~\eqref{eq:Eq2_9}, $\hat{h} \sim \sum^N_{i=1} z_i^2 \hat{n}_i^0 \sim \sum^N_{i=1} z_i^2 \hat{n}_i^\infty \sim \hat{I}$, where $\hat{I}$ is the non-dimensionalized mixture ionic strength.  Combining these relations, $\hat{M} \sim \hat{I}$.  Second, recall that $f = q / (p + q)$.  Stoichiometry requires that the ionic strength of $\text{KCl}$ is $I_{\text{KCl}} = (1/2) (z^2_{\text{K}^+} p + z^2_{\text{Cl}^-} p) \;\text{M} = p \;\text{M}$, with $z_{\text{K}^+} = -z_{\text{Cl}^-} = 1$.  The ionic strength of $\text{LaCl}_3$ is $I_{\text{LaCl}_3} = (1/2) (z^2_{\text{La}^{3+}} q + z^2_{\text{Cl}^-} q) \;\text{M} = 6q \;\text{M}$, with $z_{\text{La}^{3+}} = 3$ and $z_{\text{Cl}^-} = -1$.  Note that $I_{\text{LaCl}_3}$ increases nonlinearly with an increase in the ion valence $z^2_{\text{La}^{3+}}$.  Thus, $I = I_{\text{KCl}} + I_{\text{LaCl}_3} = p+6q \;\text{M}$.  Upon non-dimensionalization, $\hat{I}_{\text{LaCl}_3} = 6\hat{q}$ and $\hat{I} = \hat{q}(5f+1)/f$.  Recall that $\hat{M} \sim \hat{I}$, thus $\hat{M}_{\text{LaCl}_3} \sim \hat{I}_{\text{LaCl}_3}$, where $\hat{M}_{\text{LaCl}_3}$ is the contribution of $\text{LaCl}_3$ to the mobility of the mixture $\hat{M}$.  Thus, it is established that
\begin{equation}
    \label{eq:Eq3_1}
    \frac{\hat{M}_{\text{LaCl}_3}}{\hat{M}} = \frac{6f}{(1+5f)},
\end{equation}
\begin{equation}
    \label{eq:Eq3_2}
    \frac{\hat{M}_{\text{MgCl}_2}}{\hat{M}} = \frac{3f}{(1+2f)},
\end{equation}
where eqn~\eqref{eq:Eq3_2} is obtained for a $\text{KCl}/\text{MgCl}_2$ mixture following a similar procedure as eqn~\eqref{eq:Eq2_13}.  Fig.~\ref{fig:Fig5}(c) shows the nonlinear relations of $\hat{M}_{\text{LaCl}_3}/\hat{M}$ and $\hat{M}_{\text{MgCl}_2}/\hat{M}$ to $f$.  Note that $\hat{M}_{\text{LaCl}_3}/\hat{M}$ (blue line) shows a stronger increase than $\hat{M}_{\text{MgCl}_2}/\hat{M}$ (black line) at the same $f$.  This is because the nonlinearity in the ion valence for $z^2_{\text{La}^{3+}} = 9$ is stronger than that for $z^2_{\text{Mg}^{2+}} = 4$.  This explains why in Fig.~\ref{fig:Fig5}(a) and (b) the sensitivity of $\hat{M}$ to a change in $f$ in a $\text{KCl}/\text{LaCl}_3$ mixture is stronger than that in a $\text{KCl}/\text{MgCl}_2$ mixture.

\textcolor{black}{Before closing, we demonstrate the generality of the first two key findings of this work, which were demonstrated in Section~\ref{sec:Sec3_1} with a valence asymmetric electrolyte, in an electrolyte mixture.}  First, recall the first key finding that the magnitude of diffusiophoresis weakens significantly as the porous medium permeability decreases.  Fig.~\ref{fig:Fig6}(a) shows the diffusiophoretic mobility $\hat{M}$ versus the colloid surface potential $\hat{\zeta}$ in different porous media saturated with a $\text{KCl}/\text{LaCl}_3$ mixture of composition $f = 0.5$.  The first key finding is confirmed that the magnitude of $\hat{M}$ decreases substantially as $l/a$ decreases (from black to green line).  In fact, the first key finding holds not just in a mixture of two electrolytes.  Fig.~\ref{fig:Fig6}(b) shows $\hat{M}$ versus $\hat{\zeta}$ in different porous media saturated with a phosphate-buffered saline (PBS) solution, which is a mixture of $\text{NaCl}$, \text{KCl}, $\text{Na}_2\text{HPO}_4$, and $\text{KH}_2\text{PO}_4$.  The detailed ionic composition of the PBS solution is given in Appendix~A.  Again, the first key finding is confirmed by the decreasing magnitude of $\hat{M}$ with decreasing $l/a$ (from black to green line).  Lastly, recall the second key finding that, with a valence asymmetric electrolyte, diffusiophoresis in a denser porous medium can be stronger than diffusiophoresis in a less dense porous medium saturated with a valence symmetric electrolyte.  Fig.~\ref{fig:Fig6}(c) shows $\hat{M}$ versus $\hat{\zeta}$ in different porous media saturated with different $\text{KCl}/\text{LaCl}_3$ mixtures.  The black line corresponds to a denser porous medium ($l/a = 0.5$) saturated with a mixture with a higher fraction of $\text{LaCl}_3$ ($f = 1$), whereas the blue line corresponds to a less dense porous medium ($l/a = 0.75$) saturated with a mixture with a lower fraction of $\text{LaCl}_3$ ($f = 0.33$).  The second key finding is confirmed by observing that for positive $\hat{\zeta}$ the magnitude of $\hat{M}$ of the black line can indeed be larger than that of the blue line.  In sum, these results underline the novelty of this work and the importance of leveraging valence asymmetric electrolytes and electrolyte mixtures to achieve richer responses in diffusiophoresis in porous media.

\section{\label{sec:Sec4}Conclusions}
In this work, we have developed a predictive model for diffusiophoresis in porous media saturated with a general mixture of electrolytes.  Prior models have focused either on diffusiophoresis in a mixture of electrolytes in the absence of porous media,\citep{Pawar93, Majhi22, Chiang14, Shi16, Gupta19, Ohshima21, Samanta23, Bhaskar23a} or diffusiophoresis in porous media saturated with a valence symmetric electrolyte.\citep{Doan21, Sambamoorthy23, Bhaskar23b, Somasundar23}  The present model is novel in that it accounts for the coupled effects of porous media and electrolyte mixtures.  We have computed the diffusiophoretic mobilities in porous media saturated with a valence asymmetric electrolyte as well as in porous media saturated with a mixture of valence symmetric and asymmetric electrolytes.  We summarize the three key findings below.

\textcolor{black}{The first key finding of this work is that, in the same electrolyte solution, decreasing the porous medium permeability weakens the colloid diffusiophoretic motion significantly.  This is consistent with the fundamental nature of porous media which provides a larger hydrodynamic hindrance to the colloid and the suspending fluid, and thus, weakens diffusiophoresis.  This key finding is consistent with the same conclusion drawn in prior work for diffusiophoresis in porous media saturated with a valence symmetric electrolyte.\citep{Doan21, Sambamoorthy23, Bhaskar23b, Somasundar23}  Here, we have generalized this key finding to diffusiophoresis in porous media saturated with a valence asymmetric electrolyte and a general mixture of electrolytes.}

\textcolor{black}{The second key finding is that, by utilizing a valence asymmetric electrolyte, diffusiophoresis in a denser porous medium can be stronger than that in a less dense porous medium saturated with a valence symmetric electrolyte.  This is contrary to what one might expect that diffusiophoresis in a denser porous medium is always weaker.  In fact, this surprising result is due to the fact that valence asymmetric electrolyte can generate a stronger electrophoretic motion of the colloid, which outweighs the hydrodynamic drag due to the porous medium.  We have further generalized this key finding by showing that diffusiophoresis in a denser porous medium saturated with a mixture of higher fraction of asymmetric electrolyte can be stronger than that in a less dense porous medium saturated with a mixture of lower fraction of asymmetric electrolyte.}

\textcolor{black}{The third key finding is that varying the mixture composition can change not only the magnitude of the diffusiophoretic motion significantly, but also qualitatively its direction.  The diffusiophoretic mobility in an electrolyte mixture is not simply the superposition nor the weighted average of the contributions from its individual electrolyte.  This is because diffusiophoresis is governed by the nonlinearly coupled Poisson-Nernst-Planck-Brinkman equations.  This underlines the value of the present work to determine the colloid diffusiophoretic motion by solving the coupled governing equations.}  Together, the present model and these key findings will enable fundamental understanding of diffusiophoresis in porous media and predict a richer set of diffusiophoresis responses in the presence of an electrolyte mixture.

\section*{\label{sec:AppA}Appendix A: Ionic composition of phosphate-buffered saline solution}
A $1\times$ phosphate-buffered saline (PBS) solution comprises $137 \;\text{mM}$ $\text{NaCl}$, $2.7 \;\text{mM}$ $\text{KCl}$, $10 \;\text{mM}$ $\text{Na}_2\text{HPO}_4$, and $1.8 \;\text{mM}$ $\text{KH}_2\text{PO}_4$.  From dissociation, there are $157 \;\text{mM}$ $\text{Na}^+$, $139.7 \;\text{mM}$ $\text{Cl}^-$, $4.5 \;\text{mM}$ $\text{K}^+$, $10 \;\text{mM}$ $\text{HPO}_4^{2-}$, and $1.8 \;\text{mM}$ $\text{H}_2\text{PO}_4 ^-$.  Dihydrogen phosphate ion $\text{H}_2\text{PO}_4 ^-$ and monohydrogen phosphate ions $\text{HPO}_4^{2-}$ have a chemical equilibrium, $\text{H}_2\text{PO}_4 ^- \rightleftharpoons \text{H}^+ + \text{HPO}_4^{2-}$, with an equilibrium constant of $10^{-4.21}$ $\text{mM}$ and diffusivities $D_{\text{HPO}_4^{2-}} = 0.439 \times 10^{-9} \;\text{m}^2 \;\text{s}^{-1}$ and $D_{\text{H}_2\text{PO}_4^{-}} = 0.879 \times 10^{-9} \;\text{m}^2 \;\text{s}^{-1}$.\citep{Haynes14, Dane02}  Thus, there is $1.11 \times 10^{-5} \;\text{mM}$ $\text{H}^+$, which is negligible compared to the five dominant species, namely, $\text{Na}^{+}$, $\text{K}^{+}$, $\text{Cl}^{-}$, $\text{HPO}_4^{2-}$, and $\text{H}_2\text{PO}_4^{-}$.  To obtain the concentrations of the five dominant species at $I = 1 \;\text{mM}$ as in Fig.~\ref{fig:Fig6}(b), we first obtain the ratios of the concentrations of the five species to the ionic strength of the mixture as
\begin{equation}
\label{eq:EqA_1}
\begin{split}
&\quad\quad\quad\quad\quad \frac{n^{\infty}_{\text{Na}^{+}}}{I} = \frac{157}{171.5}, \quad \frac{n^{\infty}_{\text{K}^{+}}}{I} = \frac{4.5}{171.5},\\
&\frac{n^{\infty}_{\text{Cl}^{-}}}{I} = \frac{139.7}{171.5}, \quad \frac{n^{\infty}_{\text{HPO}_4^{2-}}}{I} = \frac{10}{171.5}, \quad \frac{n^{\infty}_{\text{H}_2\text{PO}_4^{-}}}{I} = \frac{1.8}{171.5}.
  \end{split}
\end{equation}
The above ratios remain unchanged for PBS solutions at other ionic strengths.  Thus, the concentrations of the five dominant species at $I = 1 \;\text{mM}$ can be obtained by multiplying the above ratios to $I = 1 \;\text{mM}$, giving $0.915 \;\text{mM}$ $\text{Na}^+$, $0.0262 \;\text{mM}$ $\text{K}^+$, $0.815 \;\text{mM}$ $\text{Cl}^-$, $0.0583 \;\text{mM}$ $\text{HPO}_4^{2-}$, and $0.0105 \;\text{mM}$ $\text{H}_2\text{PO}_4 ^-$.

\section*{Conflicts of interest}
There are no conflicts of interest to declare.

\section*{Data availability}
All data that supports this study is included in this article.

\section*{Acknowledgements}
H. C. W. Chu acknowledges the funding support from University of Florida.  This work was supported by the donors of ACS Petroleum Research Fund under Doctoral New Investigator Grant 66915-DNI9.  H.C.W.C served as Principal Investigator on ACS PRF 66915-DNI9 that provided support for S.S.  The authors acknowledge the fruitful discussion with Professor Anthony J. C. Ladd.




\renewcommand\refname{References}


\providecommand*{\mcitethebibliography}{\thebibliography}
\csname @ifundefined\endcsname{endmcitethebibliography}
{\let\endmcitethebibliography\endthebibliography}{}

\end{document}